\documentclass[aps,twocolumn,floatfix,pre]{revtex4-1}
\usepackage{latexsym,amssymb,graphicx,amsmath,epsfig,calc,times}

\usepackage[usenames]{color}

\citeindextrue

\def\btt1{{\tt$\backslash$\string1}}%

\def\AmS{{\protect\the\textfont2
        A\kern-.1667em\lower.5ex\hbox{M}\kern-.125emS}}

\def\rhotwod{\rho_\text{2d}}

\def\ps{p_\text{s}}
\def\ptwod{p_\text{2d}}
\def\Ntot{N_\text{tot}}

\def\pprot{p_\text{prot}}
\definecolor{Blue}{rgb}{0,0.0,1.0}
\definecolor{Red}{rgb}{1.0,0.0,0.0}
\definecolor{Green}{rgb}{0.0,0.35,0.0}
\definecolor{Grey}{rgb}{0.5,0.5,0.5}
\newcommand{\fff}{f_\text{tot}}
\newcommand{\kt}{k_\text{B}T}

\newcommand{\rhom}{\rho_\text{m}}

\begin{document}
\title{Theoretical calculation of the phase behavior of colloidal membranes}
\author{Yasheng Yang}
\author{Michael F. Hagan}
\affiliation{Department of Physics, Brandeis University, Waltham, MA, 02454}
\email{hagan@brandeis.edu}
\date{\today}

\begin{abstract}
We formulate a density functional theory that describes the phase behavior of hard rods and depleting polymers, as realized in recent experiments on suspensions of \emph{fd} virus and non-adsorbing polymer. The theory predicts the relative stability of nematic droplets, stacked smectic columns, and a recently discovered phase of isolated monolayers of rods, or colloidal membranes. We find that a minimum rod aspect ratio is required for stability of colloidal membranes and that collective protrusion undulations are the dominant effect that stabilizes this phase. The theoretical predictions are shown to be qualitatively consistent with experimental and computational results.
\end{abstract}
\maketitle

\section{Introduction}
\label{sec:intro}
The relationship between intermolecular interactions and structure is fundamental to statistical mechanics and materials science. Hard particles, which interact solely by steep repulsive potentials that prohibit overlap, have served as essential model systems for understanding this relationship. Studies of hard spheres elucidated the structure of liquids \cite{Weeks1971} and 3D crystalline phases \cite{chandler1983,Pusey1986} and investigations of hard rods demonstrated the existence of 3D nematic and smectic phases \cite{Onsager1949,Frenkel1988,Bolhuis1997}. Since all accessible hard particle configurations have no interparticle interaction energy, these results showed that entropy alone can drive the self-assembly of structures with long-range order. While the phase diagram of purely repulsive hard rods is well known~\cite{Bolhuis1997}, adding non-adsorbing polymers introduces attractive interactions between rods through the depletion effect, which leads to myriad novel equilibrium phases and metastable morphologies that are poorly understood~\cite{Asakura1954,Adams1998,Dogic2001,Dogic2003}.
Of particular interest from a structural perspective, recent experiments on suspensions of monodisperse rod-like colloidal viruses and the non-adsorbing polymer Dextran demonstrated assembly of `colloidal membranes' comprised of a one rod-length thick monolayer of colloidal rods ~\cite{Barry2010}. This observation has fundamental and practical significance. Unlike other examples of entropy-driven assembly which lead to long-range order in three dimensions, the colloidal membranes are self-limited to the thickness of a single rod in one dimension and thus are 2D structures. From a practical perspective, equilibrium colloidal membranes could enable manufacture of inexpensive and easily scalable optoelectronic devices~\cite{Baker2010}.

Previous approaches towards assembly of colloidal membranes employed chemically heterogeneous rods that mimic the amphiphilic nature of lipids which comprise  biological membranes~\cite{Park2004}. In contrast,  the {\it fd} molecules involved in assembly of colloidal membranes are structurally homogeneous, suggesting that geometry as well as chemical heterogeneity can be used to design molecules that assemble into particular structures. Doing so, however, requires a fundamental understanding of the forces that conspire to self-limit assembly and the relationship between molecular design parameters and equilibrium structures. In this article, we therefore  construct a theoretical description of colloidal rods in the presence of depletant molecules. The theory is built on insights from recent experiments on suspensions of {\it fd} virus in Dextran \cite{Barry2010} and simulations of hard spherocylinders in depletant \cite{Yang2011}. We derive a
density functional theory for a system of hard cylinders and depletant. In particular, we use an equation of state for hard disks in two dimensions to calculate the equilibrium areal density of rods within the membrane, apply free volume theory \cite{Lekkerkerker1992} to calculate the volume that the membrane excludes to polymers, and use a virial expansion to calculate rod-rod interactions between nearby membranes.  The calculations yield predictions for the relationships between osmotic pressure, rod aspect ratio, membrane properties, and phase behavior. These expressions are derived under the simplifying assumption that rod orientations are parallel to a fixed axis, and are extensively compared to computational and experimental results. We find that the theory successfully describes the interplay between configurational entropy of rods within membranes, depletion interactions, and the resulting phase behavior of colloidal rods in depletant. We demonstrate that the dominant effect which stabilizes isolated membranes arises from the entropic penalty associated with suppression of protrusion fluctuations of rods within membranes when the membranes stack. While this effect was originally modeled \cite{Barry2010} based on theory describing protrusions of individual rods \cite{Israelachvili1992}, we find that repulsions are driven primarily by collective protrusion undulations, consistent with simulations \cite{Yang2011}.

This paper is organized as follows. We first review the experiments and simulations on hard rods in the presence of depletant molecules in section \ref{sec:experiments}. We construct a simplified theoretical model for hard rods in depleting polymers and associated free energy as a functional of the rod density distribution in section \ref{sec:theoreticalModel}.  We then use the theory to predict the optimal density distributions of rods and hence the system phase behavior in section \ref{sec:results}. Under conditions for which membranes are stable, we analyze the distribution of rods within an isolated membrane. We then examine how this distribution changes as two membranes approach each other along the membrane normals, and thereby evaluate the coupling between attractive depletion interactions and  repulsive protrusion interactions as a function of membrane separation. Finally, predictions for protrusion distributions in isolated and interacting membranes, the total interaction potential between membranes, and the system phase behavior are compared to results of molecular simulations of  a rod-depletant suspension \cite{Yang2011}. We find that agreement between the theory and simulations is quantitative  for the nematic/colloidal membrane coexistence osmotic pressure and qualitative for the colloidal membrane/smectic-like stacks coexistence pressure.

\section{Previous experimental studies of colloidal membranes}
\label{sec:experiments}
\begin{figure}
\epsfxsize=0.79\columnwidth\epsfbox{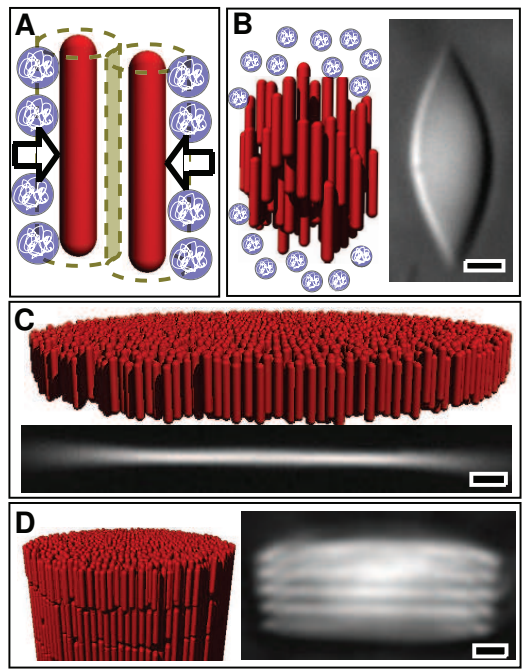}
\caption{Schematic illustrations and optical micrographs of the self-assembled structures observed in suspensions of the filamentous virus {\it fd} and non-adsorbing polymer~\cite{Barry2010}. {\bf A)} Non-adsorbing polymer induces effective attractive interactions between rods. {\bf B)} DIC micrograph and schematic of a nematic tactoid formed at low depletant concentration. {\bf C)} At intermediate depletant concentrations, rod-like viruses condense into macroscopic one rod-length 2D fluid-like membranes. {\bf D)} At high depeltant concentration, membranes stack on top of one another, forming smectic filaments. All scale bars are 5$\mu$m.
\label{fig:experiment} }
\end{figure}

We first review recent experiments on suspensions of monodisperse rod-like colloidal viruses and the non-adsorbing polymer Dextran~\cite{Barry2010} (Fig.~\ref{fig:experiment}), which motivate our theory. \emph{fd} viruses alone approximate the behavior of homogenous rods interacting with repulsive hard-core interactions~\cite{Purdy2003}. The polymer induces an entropy-driven attractive (depletion) potential between the rods, the strength and range of which can be tuned by changing the polymer concentration and radius of gyration respectively (Fig.\ref{fig:experiment}A)~\cite{Asakura1954}.
 At high polymer concentrations (attraction strength) dilute viruses condense into smectic-like stacks of 2D membranes (Fig.\ref{fig:experiment}D)~\cite{Frenkel2002}. Below a threshold polymer concentration, individual 2D monolayers (membranes) within a smectic filament unbind, indicating that the membrane-membrane interaction switches from attractive to repulsive~\cite{Barry2010} (Fig.\ref{fig:experiment}C). The monolayer membranes are stable over months or longer and can be many millimeters in diameter. As the polymer concentration is decreased further past a second threshold, membranes become unstable to nematic liquid crystalline droplets or tactoids (\ref{fig:experiment}B)~\cite{Dogic2001}. While the properties of tactoids and configurations of rods within them have been explored computationally (e.g. \cite{Trukhina2008,Trukhina2009}) and theoretically (e.g. \cite{Kaznacheev2002,Kaznacheev2003,Prinsen2003,Prinsen2004,Prinsen2004a}), theoretical models of 2D colloidal membranes are lacking.

 Although experiments conclusively demonstrated the existence of colloidal membranes, the molecular mechanisms that control their stability remained unclear. The depletion interaction that drives lateral association of rods also generates an attractive interaction between vertically adjacent membranes. For isolated membranes to be stable against stacking, there must be a repulsive interaction that overwhelms the depletion interaction.
Experiments on colloidal membranes containing a low volume fraction of fluorescently labeled rods revealed significant protrusions of rods from isolated membranes, the magnitude of which could be tuned by changing the concentration of non-adsorbing polymer. In comparison, these protrusion fluctuations were suppressed in stacked membranes~\cite{Barry2010}.
 Based on that observation, a model was proposed in which the entropy penalty associated with suppressing protrusion fluctuations of individual rods ~\cite{Israelachvili1992} leads to repulsive interactions membrane-membrane interactions under moderate osmotic pressure. However, other plausible factors could also lead to the observation of isolated membranes, including attractive interactions between virus tips and depletant, repulsions due to bending (Helfrich) modes, or kinetic trapping of metastable membrane intermediates. In Ref. \cite{Yang2011} we used computer simulations (described in appendix \ref{sec:simulations} to show that collective protrusion effects alone are sufficient to produce qualitative agreement with the experimental observations. The simulations predicted that there is a minimum aspect ratio below which colloidal membranes are never stable; this prediction was confirmed by further experiments \cite{Yang2011}. Here we develop a theory with which we  explore the origins of the stabilization of colloidal membranes in greater detail.

\section{Theoretical model}
\label{sec:theoreticalModel}
In this section we use insights obtained from our previous simulations of colloidal membranes \cite{Yang2011} to derive a tractable theoretical model for colloidal membranes.
Because the simulations suggest that large rod aspect ratios are essential for the existence of stable membranes, we cannot model colloidal membranes by directly applying previous theoretical approaches used for studying bulk phases of rod-polymer or rod-sphere mixtures \cite{Warren1994,Martinez-Raton2006,Tuinier2007,Chen2004,Chen2002}.

Following the simulation model \cite{Yang2011}, we consider hard rods and polymers that can freely interpenetrate other polymers but act  as hard particles when interacting with rods \cite{Asakura1954}. For simplicity, rods are represented as cylinders with diameter $\sigma$ and length $L$, and polymers are represented as short cylinders with diameter $\delta$ and height $h=\frac{2}{3}\delta$. The parameter
$h$ is defined such that a polymer cylinder has the same volume as a sphere of diameter $\delta$ (the traditional theoretical representation of a depletent). Cylinders can interpenetrate one another but experience hard-core interactions with rods. Note that because we consider an ideal osmolyte (ghost cylinders) we do not observe alternating layers of rods and depletents, which were described for a model with hard-sphere depletents ~\cite{Koda1996}.

We focus on conditions relevant to the experiments, where rods have large aspect ratios and are immiscible with polymers \cite{Barry2010}. We showed previously \cite{Yang2011} that under these conditions membrane bending modes,  which involve deviations of rod orientations from the membrane normal ~\cite{Helfrich1984}, are high-energy in comparison to protrusions of rods from the membrane surface ~\cite{Goetz1999} on length scales that control membrane-membrane stacking. Bending modes can  thus can be neglected when evaluating phase behavior, and  we simplify our calculation by constraining rod orientations to be parallel to a fixed direction (the $\hat{z}$ axis).

We use the density of rods with center of mass at position $\vec r$, $\rho(\vec r)$, to describe system configurations. To investigate macroscopic membranes we consider periodic boundary conditions in the $xy$-plane and assume that the density depends only on $z$, $\rho(\vec r)=\rho(z)$. The latter simplification follows from constraining rod orientations  parallel to $\hat{z}$. Then a peak in $\rho(z)$ corresponds to a membrane which is macroscopic in two dimensions (e.g. Fig.~\ref{fig:histZSingleLayer} below). The width of the peak reflects the size of the membrane in the $z$-direction and thus the extent of the protrusion distribution.


{\bf Free volume theory for the free energy.}
 We describe rod-rod interactions with a third order virial expansion and  rod-polymer interactions with the free volume approach presented in Ref.~\cite{Lekkerkerker1992}, adapted to describe the 2-D cross-sections of membranes. The results are close to those of a complete third order virial expansion, which is lengthy and is not presented here.

In the free volume theory \cite{Lekkerkerker1992}, for a particular rod density distribution  $\rho(z)$ the free energy per unit area, $\fff$, can be written as

\begin{align}
S_{xy}\beta \fff=&\int d\mathbf 1 \rho(\mathbf 1) (\ln \rho(\mathbf 1)\lambda^3 - 1)\nonumber\\
&-\frac{1}{2}\int d\mathbf 1 d\mathbf 2 \rho(\mathbf 1) \rho(\mathbf 2) f(\mathbf 1,\mathbf 2) \nonumber\\
&-\frac{1}{6}\int d\mathbf 1 d\mathbf 2 d\mathbf 3
\rho(\mathbf 1)\rho(\mathbf 2)\rho(\mathbf 3)f(\mathbf 1,\mathbf 2)f(\mathbf 1,\mathbf 3)f(\mathbf 2,\mathbf 3) \nonumber\\
&+S_{xy}\beta \ps\int dz(1-\alpha(z))
\label{eq:freeEnergyOriginal}
\end{align}
where $S_{xy}\equiv \int dx dy$ is the total area of the membrane, bold numbers are the spatial coordinates,
and $f(\mathbf 1,\mathbf 2)$ is the Mayer function between rods.
In Eq.~\ref{eq:freeEnergyOriginal}, the first term is the ideal gas free energy and
the following two terms are respectively the second and third order virial terms for rod-rod interactions. The second virial term represents the pairwise mutual excluded volume interaction between a rod and its neighbors, and the third virial term accounts for the mutual exclusion among three rods. In the numerical minimization of free energy of multi-membranes described below, we found that a free energy expression with only the second order virial term often leads to merging of membranes and unphysically high rod densities, thus the three body effects in the third order virial term are necessary for physical results. This is to be expected, since  a second order virial expansion is inaccurate for parallel rods \cite{Onsager1949}.

\begin{figure}
\epsfxsize=0.75\columnwidth\epsfbox{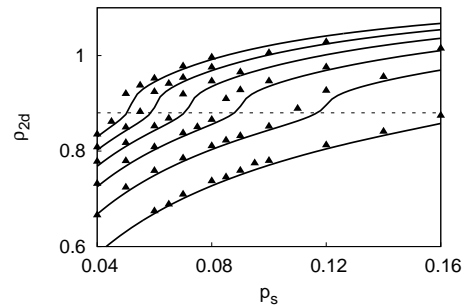}
\caption{Surface density of rods in isolated membranes. Symbols are the results of simulations (appendix \ref{sec:simulations}). Lines are predictions calculated from the equation of state for a 2-dimensional hard disk system\cite{Luding2001} as described in the text. The rod lengths, from top to bottom, are $L=175$, $150$, $125$, $100$, $75$ and $50$. The diameter of a polymer sphere in the simulations is $\delta=1.5$ and thus a diameter $h=1$  is used for the theoretical cylinders. The dashed line indicates the freezing density of a hard disk system, $\rho=0.88$\cite{Lowen1993,Truskett1998}.
 \label{fig:rho2d}
}
\end{figure}

 The last term in Eq.~\ref{eq:freeEnergyOriginal} is the free energy due to the volume that rods exclude to spheres. The variable $\alpha(z)$ describes the free area available to polymers at position $z$, and depends on rod densities at any center of mass position from which rods can overlap. Specifically, the total density of rods that could overlap with a cylinder of height $h$ at $z$ is $\rho^*(z) = \int_{z-L/2-h/2}^{z+L/2+h/2}dz\rho(z)$. If we assume that the $xy$ distribution of rods is not perturbed by polymer,  then the fraction of free area $\alpha(z)$ can be calculated from $\rho^*(z)$,
using the result of scaled particle theory for two-dimensions\cite{Lebowitz1965},
\begin{align}
\alpha(z) = (1-\phi)\exp(-2\gamma\eta-\gamma\eta^2-\gamma^2\eta^2)
\end{align}
with $\phi\equiv\pi \sigma^2\rho^*(z)/4$, $\gamma\equiv\phi/(1-\phi)$, and $\eta \equiv \delta/\sigma$.
Finally the excluded volume per unit area for the whole membrane is $\int dz(1-\alpha(z))$.
By using the scaled particle theory, the effects of overlapping excluded volumes of protruding rods are considered approximately, and
hence the rod-rod correlations are partially included. As mentioned above, this approximation is equivalent to a third order virial expansion of rod-polymer interaction.

Note that the integration over the $z$ coordinate should be restricted to a finite region to avoid divergence, since there is always an arbitrarily low but finite concentration of rods in the solution, and we will do so in the following numerical calculations.

Since the rod density depends only on $z$, integration over the $x$ and $y$ directions can be carried out analytically to give the final form for the free energy as
\begin{align}
\beta \fff=&\int dz \rho(z)(\ln\rho(z) -1) \nonumber\\
&+\frac{1}{2}A\int dz_1 \rho(z_1)(\rho^\dagger(z_1+L/2)+\rho^\dagger(z_1-L/2))\nonumber\\
&+\frac{1}{3}B\int dz_1 \rho(z_1)\int_{z_1-L}^{z_1}dz_2\rho(z_2)\rho^\dagger(z_2+L/2)\nonumber\\
&+\frac{1}{6}B\int dz_1 \rho(z_1)(\rho^\dagger(z_1+L/2))^2\nonumber\\
&+ \beta \ps\int dz(1-\alpha(z))
\label{eq:freeEnergy}
\end{align}
with constants $A$ and $B$ given in appendix ~\ref{sec:integration}, and the cumulative density defined as
\begin{align}
\rho^\dagger(z) \equiv \int_{z-L/2}^{z+L/2}dz\rho(z)
\label{eq:rhoDagger}
\end{align}
The details of the calculation are provided in appendix ~\ref{sec:integration}.

The equilibrium rod distribution can be acquired by minimizing $\fff$ with respect to $\rho(z)$. We will see that Eq.~\ref{eq:freeEnergy} can qualitatively describe features of the phase behavior and membrane-membrane interactions. However, it does not give an accurate prediction of the areal density of rods within the membrane, $\rhotwod$. This limitation is to be expected, since areal densities are high and membranes are even crystallized at high osmotic pressures according to the simulations.
 To overcome this limitation, we independently relate $\rhotwod$ to the osmotic pressure $\ps$, the rod length $L$ and the polymer size $h$ using the equation of state for 2D hard disks \cite{Luding2001}, modified to account for the extent of the rods in the $z$ direction. The adaptation of the equation of state to rods within the membrane is described in appendix ~\ref{sec:rho2dTheory}. The equation of state relating $\rhotwod$ to a 2D pressure $\ptwod$ is given in Eqs. \ref{eq:state2d} to \ref{eq:Cfour} and the 2D pressure is given by $\ptwod\approx (L+h)\ps$. The predicted areal densities are compared to simulation results in Fig.~\ref{fig:rho2d}.

The equation of state value of the areal density is used as a constraint when minimizing the free energy,
\begin{align}
0=C\equiv \int dz\rho(z) - m \rho_\text{2d}(L,h,\ps)
\label{eq:constraint}.
\end{align}
Here the quantity $m$ fixes the total number of rods in the system $\Ntot$ as $\Ntot=m \rhotwod S_{xy}$. Under conditions for which isolated or stacked membranes are stable with respect to the nematic phase, $m$ will correspond to the number of membranes in the system.

 Minimizing $\fff$ thus requires
\begin{align}
0=\frac{\delta \beta \fff}{\delta \rho(z)} - \zeta \frac{\delta C}{\delta \rho(z)}
\label{eq:fmin}
\end{align}
with $\zeta$ a Lagrange multiplier. Substituting Eqs.~\ref{eq:freeEnergy} and \ref{eq:constraint} into Eq.~\ref{eq:fmin} then results in an integral equation for the rod distribution
\begin{align}
\rho(z)=&\exp(-\zeta)\exp(\beta \ps\int dz_1\frac{\partial\alpha(z_1)}{\partial\rho(z)}) \nonumber\\
	&\times \exp(-A(\rho^\dagger(z+L/2)+\rho^\dagger(z-L/2)))\nonumber\\
	&\times \exp(-B\int_{z-L}^{z}dz_1\rho(z_1)\rho^\dagger(z_1+L/2)) \nonumber\\
	&\times \exp( - \frac{1}{2}B(\rho^\dagger(z+L/2))^2)
\label{eq:multiLayerRho}
\end{align}
The detailed derivation of Eq.~\ref{eq:multiLayerRho} is given appendix \ref{sec:differentiation}.

Eq.~\ref{eq:multiLayerRho} along with Eq.~\ref{eq:constraint} can be solved numerically to obtain the equilibrium rod distribution  $\rho(z)$ for a specified value of $m$.

\section{Theory results}
\label{sec:results}

\label{sec:phaseDiagramTheory}
 In this section we analyze the behavior predicted by Eqs.~\ref{eq:multiLayerRho} and ~\ref{eq:constraint}. For low and moderate osmotic pressures $\ps$, we will see that stacks of membranes are thermodynamically unstable; either isolated membranes or nematic configurations (with no membranes) are thermodynamically stable.  Under these conditions one can set the number of membranes in Eq.~\ref{eq:constraint} to $m=1$ without loss of generality. \footnote{Under the condition $m=1$ the minimization can be simplified. Since the rod-rod interaction interaction virial terms within a single membrane are already accounted for by the areal density constraint, Eq.\ref{eq:constraint}, little numerical difference is incurred by omitting the virial terms. The virial terms must be kept for $m\ge2$ because they include the protrusion interaction.}

\subsection{Nematic/isolated membrane phase boundary.}
\label{sec:nematicIsolated}
\begin{figure}
\epsfxsize=0.75\columnwidth\epsfbox{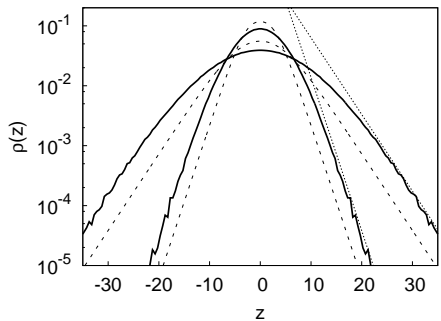}
\caption{Protrusion distribution in a single membrane. $z$ is the displacement of rods from the center of the membrane. Curves are for $\delta=1.5$, $L=100$ and $\ps=0.06$ (outer lines) or $0.12$ (inner lines).
Solid lines are the simulation results. The dashed lines are the distributions predicted by Eq.\ref{eq:multiLayerRho}, the solid lines are simulation results, and the dotted lines show the scaling expected from an analysis based on independent rod protrusions \cite{Israelachvili1992}:  $\rho(z)\sim \exp(-\ps Az)$, with $A=\pi(\sigma+\delta)^2/4$.}
 \label{fig:histZSingleLayer}
 \end{figure}

\begin{figure}[t]
\epsfxsize=0.75\columnwidth\epsfbox{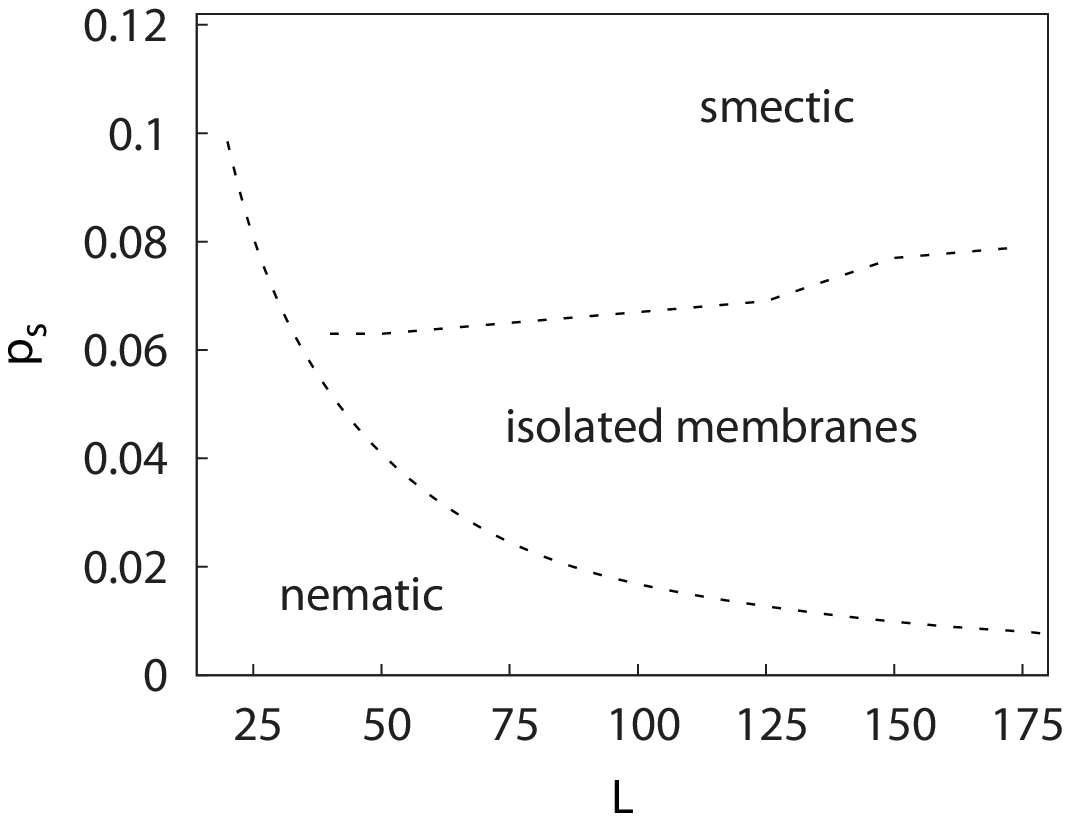}
\epsfxsize=0.75\columnwidth\epsfbox{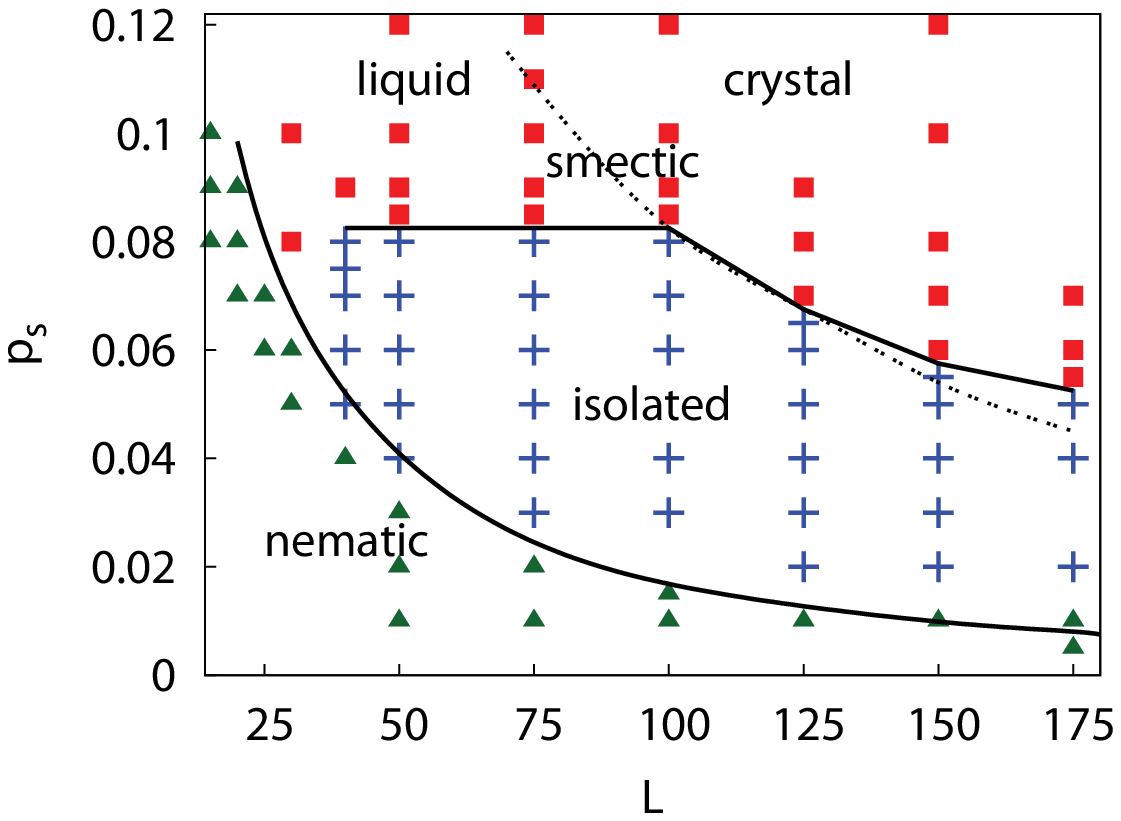}
\caption{{\bf (top)} Theoretical phase diagram as a function of osmotic pressure $\ps$ and aspect ratio $L$. The theoretical phase boundaries, calculated as described in the text, are shown as dashed lines. The depletant size is $\delta=1.5$. {\bf (bottom)} Phase behavior predicted by simulations for the same parameters. Triangles $\blacktriangle$ denote denote parameters that lead to nematic configurations, + symbols correspond to isolated membranes, and $\blacksquare$ symbols correspond to smectic layers. The lower solid line is the theoretical prediction for the nematic/colloidal membrane phase boundary, while the upper solid line is fit by eye to the  the colloidal membrane/smectic phase boundary. The dashed line indicates parameter values above which rods crystallized within simulated colloidal membranes. Simulation data is from Ref. \cite{Yang2011}.
\label{fig:phasesTheory}
}
\end{figure}

For low $\ps$, a flat distribution $\rho(z)=\mathrm{constant}$ is the unique solution to Eq.~\ref{eq:multiLayerRho}, indicating that the nematic state is the thermodynamic equilibrium. (Note that we cannot consider the isotropic to nematic transition, which would occur at lower $\ps$, because we have assumed that rods are parallels to the $z$ axis.) As $\ps$ increases past a threshold value a stable inhomogeneous solution also appears, with a peak in $\rho(z)$ that corresponds to the center of a membrane (Fig.~\ref{fig:histZSingleLayer}). This solution corresponds to an isolated colloidal membrane. The free energies of the two solutions are compared to determine the equilibrium state. As shown in Fig.~\ref{fig:phasesTheory}, the predicted coexistence curve for the nematic phase and isolated membranes shows remarkable agreement with simulation results for the rod lengths considered. As the osmotic pressure increases across the spinodal, only the inhomogeneous solution (corresponding to an isolated membrane) remains stable.

{\bf Protrusion distribution.}
We can further  investigate the ability of the theoretical model to describe colloidal membranes by comparing theoretical predictions for the distribution of rods within membranes to those measured from simulations and a mean field estimate ~\cite{Israelachvili1992}. In the latter approach the protrusion of a single rod from a membrane exposed to depletant osmotic pressure $\ps$ incurs a free energy $f_\text{rod}=\ps A d$, with $A$ the cross-sectional area of the rod and $z$ the protrusion distance. For uncorrelated protrusion sites the distribution of protrusions obeys an exponential distribution $\pprot(z) \sim \exp(-\ps A z/\kt)$, with $\pprot(z)$ the density of rods with ends located a distance $z$ above the mean surface of the membrane.
Examples of the numerical solutions for $\rho(z)$ are compared to rod distributions measured in simulations of isolated membranes and the mean field estimate in Fig.\ref{fig:histZSingleLayer}.
We see that
Eq.\ref{eq:multiLayerRho} correctly reproduces the exponential distribution at large $|z|$ (predicted by the mean field estimate)
and the broadening of the distribution at small $|z|$. The theory is more accurate than the simple mean field estimate \cite{Israelachvili1992} in this context because the free area accounts for rod-rod correlations induced by the overlapping excluded volumes of protruding neighbors.

Although the agreement between theory and simulations is good, the predicted distributions are narrower then the simulation results.
We also used the third order virial expansion (instead of the free volume approach) to calculate the rod-polymer interaction, and obtained similar results.
The fact that the theory predicts a narrower rod distribution reflects the fact that rod-rod correlations are not completely accounted for by either the free volume or third order virial expansion calculations. The quantitative accuracy likely could be improved by going to fourth order in the virial expansion, since graphs involving polymers (cylinders) that are in the excluded volume regions of two neighboring but non-overlapping rods appear at this order. We also note though that collective protrusions introduce long wavelength modes to the membrane, as shown by the height-height correlation spectrum (flicker spectrum) in Fig.~7 of Ref. \cite{Yang2011}. One can potentially account for these long wavelength modes using renormalization theory, as in Ref.~\cite{Lipowsky1993}.

\subsection{Isolated membrane/smectic phase boundary.}
\begin{figure}
\epsfxsize=0.75\columnwidth\epsfbox{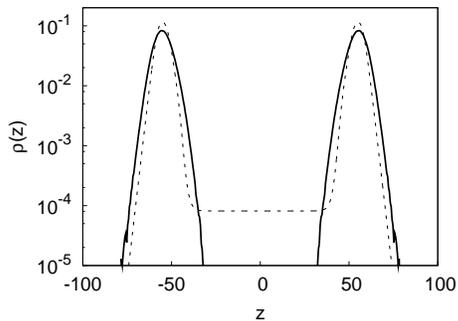}
\caption{ Rod density $\rho(z)$ of two attractive membranes. The the dashed line is the prediction of Eq.~\ref{eq:multiLayerRho} and the solid line is from simulations. Note that the two peaks are separated by approximately the rod length and thus the configuration contains two closely stacked membranes.
The parameter values are $\delta=1.5$, $L=100$, and $\ps=0.12$.
}
 \label{fig:histZDoubleLayer}
\end{figure}

For osmotic pressures at which membranes are favorable, we determine whether isolated colloidal membranes or smectic-like stacks are the thermodynamic minimum by evaluating the rod distribution for the case of two membranes by solving Eq.~\ref{eq:multiLayerRho} under the constraint of Eq.~\ref{eq:constraint} with $m=2$. Numerical solution of Eq.~\ref{eq:multiLayerRho} at osmotic pressures above the nematic-isolated phase boundary yields a stable solution for $\rho(z)$ with two peaks corresponding to two membranes. The distance between the peaks depends on the osmotic pressure and closely matches simulation results, as shown by the configuration for $\ps=0.12$ and $L=100$, for which smectic-like stacks are thermodynamically favorable, in
Fig.~\ref{fig:histZDoubleLayer}. However, there is a finite predicted rod density between the two membranes, which is likely due to truncating the virial expansion at third order, and the predicted membrane widths are narrower than those of the simulation, as discussed above.

Below a certain value of the osmotic pressure,  membrane-membrane
interactions switch from attractive to repulsive, as signified by a switch
from adjacent to separated peaks in the optimal density distribution. However, the exact pressure at which this occurs is sensitive to numerical
error and depends on the region integrated over (which sets the concentration of membranes). As noted in Ref.~\cite{Yang2011} the osmotic  pressure at which smectic-like stacks become stable must depend on the  concentration of membranes, since the membrane-membrane  interaction free energy must be sufficiently attractive to compensate  the  reduction in membrane translational entropy associated with stacking. Therefore, to accurately  predict the colloidal membranes/smectic phase coexistence osmotic pressure, we  next use the theoretical model to calculate the interaction free energy between two membranes.

{\bf Membrane-membrane interactions.}
To calculate the free energy $f(d)$ as a function of the distance between membrane centers $d$ we use an approach analogous to umbrella sampling \cite{Frenkel2002} (this approach is much simpler than performing the projection by analytical integration). We augment Eq. ~\ref{eq:fmin} with an additional constraint on $d$
\begin{align}
\frac{d}{2}=\frac{\int_{z>0}dzz\rho(z)}{\int_{z>0}dz\rho(z)}=-\frac{\int_{z<0}dzz\rho(z)}{\int_{z<0}dz\rho(z)}
\label{eq:constraintDistance}.
\end{align}
Following the umbrella sampling procedure \cite{Frenkel2002}, we implement this constraint as a penalty to the free energy:
\begin{align}
\beta f_\text{penalty}=&k\left(\frac{\int_{z>0}dzz\rho(z)}{\int_{z>0}dz\rho(z)}-\frac{d}{2}\right)^2\nonumber\\
	 &+k\left(\frac{\int_{z<0}dzz\rho(z)}{\int_{z<0}dz\rho(z)}+\frac{d}{2}\right)^2
\end{align}
with $k>0$ an adjustable constant.
We then numerically minimize $\fff+f_\text{penalty}$ to obtain the optimal density distribution $\rho(z;d)$ under the constraint Eq.~\ref{eq:constraintDistance}. Finally, the interaction free energy $f(d)$ is obtained by subtracting the penalty term: $f(d)=\fff(\rho(z;d))/m\rhotwod$.

\begin{figure}
\epsfxsize=0.49\columnwidth\epsfbox{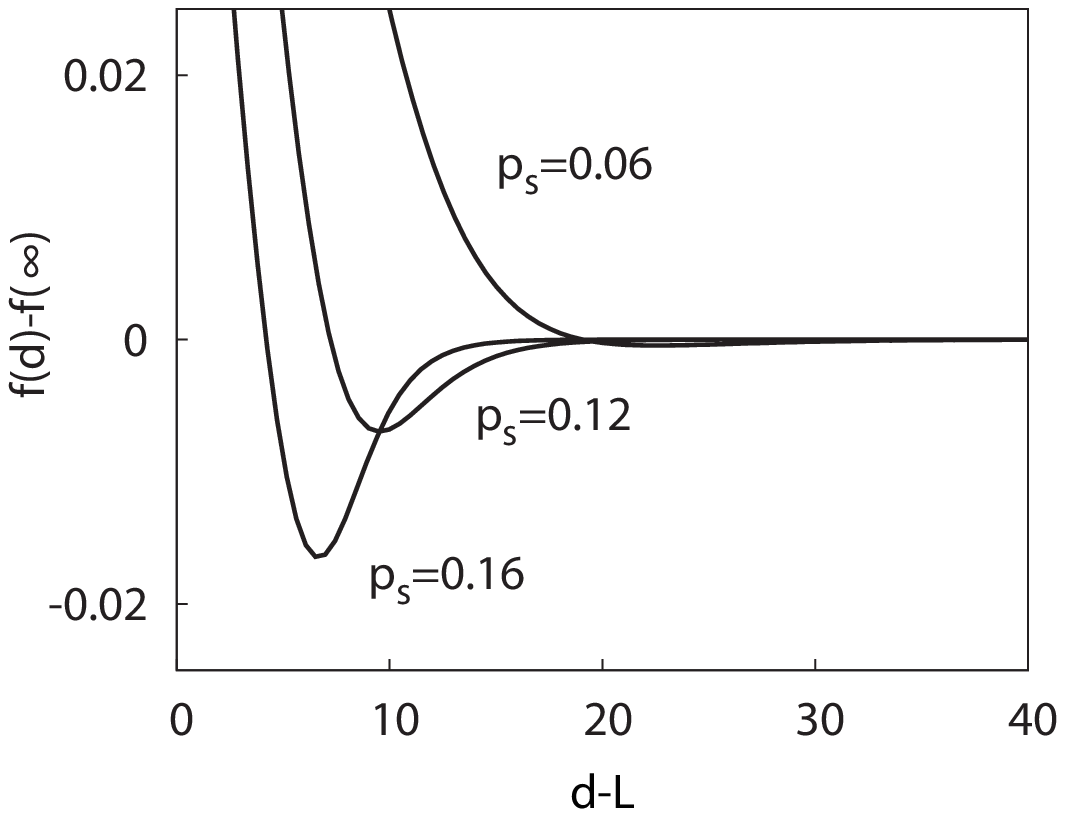}
\epsfxsize=0.49\columnwidth\epsfbox{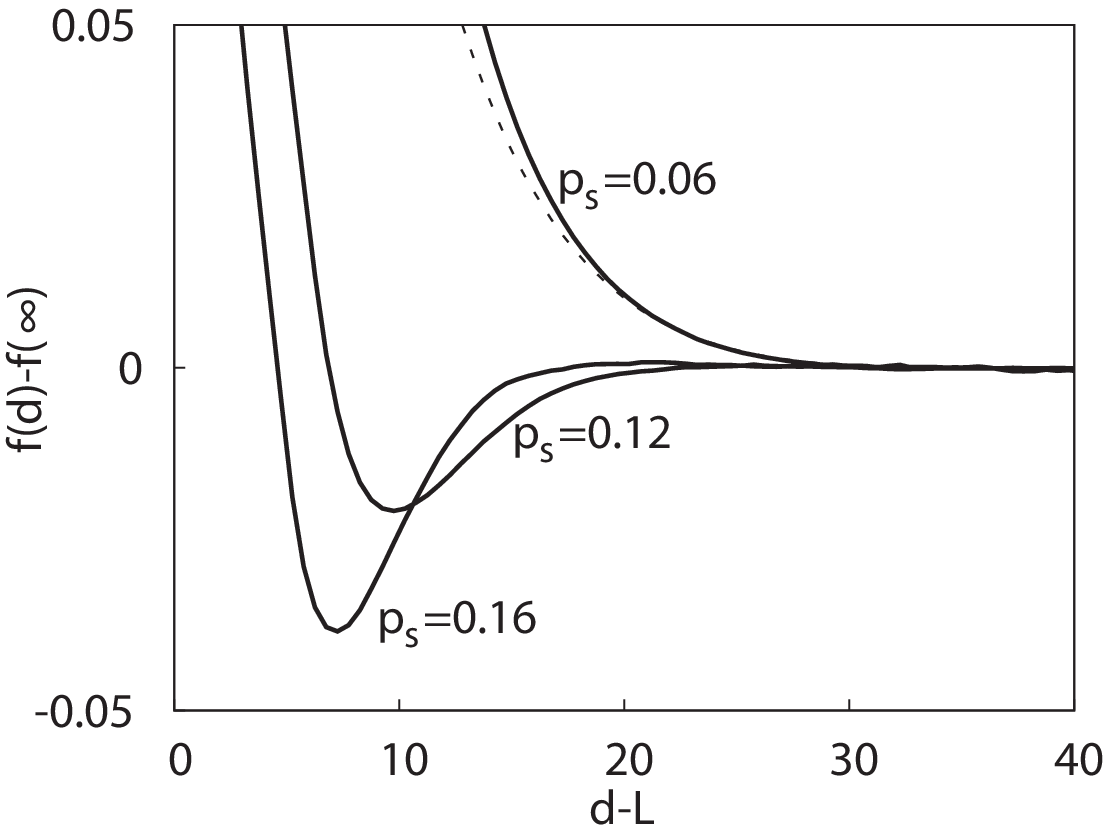}
\caption{{\bf (left)} Theoretical free energy of interacting membranes predicted by the free energy Eq.~\ref{eq:freeEnergy}
with the constraints Eq.~\ref{eq:constraint} and Eq.~\ref{eq:constraintDistance}.
Curves are for parameters $\delta=1.5$ and $L=100$, with indicated values of the osmotic pressures $\ps$. {\bf (right)} Free energy of interacting membranes from simulations using umbrella sampling at the same parameters. The solid lines correspond to calculations in which rods are constrained parallel to the membrane normals, while the dashed line at $\ps=0.06$ corresponds to a calculation in which this constraint is relaxed. Data is from Ref. \cite{Yang2011}.
\label{fig:freeEnergy}
}
\end{figure}

Examples of interaction free energies are shown in Fig.~\ref{fig:freeEnergy} (top).  If we compare  these theoretical predictions to simulation umbrella sampling results \cite{Yang2011} in Fig.~\ref{fig:freeEnergy} (bottom), we see that the theoretical $f(d)$ curves qualitatively agree with simulation results, but that the theoretical calculations have shallower attractive basins. Like the discrepancy between the theoretical and computational protrusion distributions, this quantitative difference may be due to the fact that the rod-rod correlations are not fully accounted for.

\begin{figure}
\epsfxsize=0.75\columnwidth\epsfbox{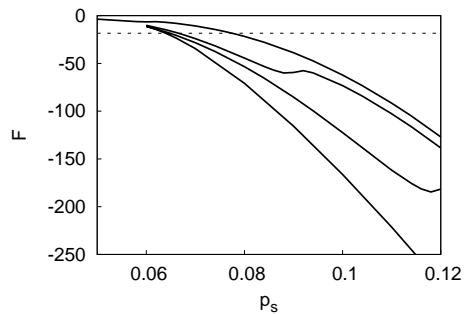}
\caption{
Theoretical calculation of the total free energy of attractive basin $F$, defined by Eq.~\ref{eq:basinF}
with $M=10^4$, for rod lengths $L=40$, $75$, $100$ and $150$ (solid lines, from left to right). The horizontal dashed line is $F_0$, defined by Eq.~\ref{eq:criterion} with $\rho_\text{m}v_0=10^{-8}$.
\label{fig:fIntgTheory}
}
\end{figure}

Following Ref.~\cite{Yang2011}, smectic layers are thermodynamically favorable at finite membrane concentration $\rhom$ if the total free energy of the attractive basin in the membrane-membrane interaction potential satisfies
 \begin{equation}
F \le F_0=\kt \ln\rho_\text{m} v_0
\label{eq:basinF}
\end{equation}
with
 \begin{equation}
 \exp(-\beta F)=\int_{f(s)<0}ds\exp(-2\beta Mf(s))
 \label{eq:criterion}
 \end{equation}
 with $M$ the number of rods in one membrane, and $v_0$ a standard state volume. We roughly estimate $M=10^4$ and $\rhom v_0 = 10^{-8}$ from the experimental conditions; the location of the phase boundary is not sensitive to the value of $\rhom v_0$.
Fig.~\ref{fig:fIntgTheory} shows the theoretical values for the free energy of attractive basins $F$ at a number of rod lengths. As expected, $F$ becomes more favorable as the osmotic pressure $\ps$ increases. The theoretical $F$ curves cross $\kt \ln\rho_\text{m} v_0$ near $\ps\sim 0.07$, which is close to the simulation results for depletant size $\delta=1.5$.

As shown in Fig.~\ref{fig:phasesTheory}, the theoretical isolated-smectic phase boundary shows reasonable agreement with the simulated phase boundary for $L\lesssim100$. Furthermore, both methods predict a similar threshold value of the aspect ratio, $L\approx30$, below which the system transitions directly from nematic configurations  to smectic-like stacks of membranes. This prediction was confirmed by experiments\cite{Yang2011} in which the osmotic pressure was varied by controlling the concentration of non-adsorbing polymer and the depletant size was varied by changing polymer radius of gyration.

Notably, the theory does not reproduce the decrease in the transition osmotic pressure at large rod lengths seen in the simulations. As described in Ref.~\cite{Yang2011}, this trend results from crystallization of rods within membranes in the simulations. Under the simulation conditions, membrane crystallization decreases the interaction free energy due to protrusions and thus lowers the isolated-smectic coexistence osmotic pressure. In contrast, the theoretical interaction free energy $f(d)$ assumes a disordered distribution of rod positions within the plane of the membrane and thus does not allow for membrane crystallization. It is worth noting that simulations in which the constraint on parallel rod orientations was relaxed required larger rod aspect ratios and/or higher osmotic pressures for crystallization of rods within membranes. Given that observation and the fact that theory and simulation agree for $L\lesssim100$,  we chose not to extend the theoretical model to allow for crystallization.

\section{Conclusions and outlook}
\label{sec:conclusions}
In summary, we have presented a theoretical model that represents hard rods in the presence of depletant molecules such as non-adsorbing polymer.
The free energy is constructed by using the free volume theory for depletant-rod interactions,
and a third order virial expansion for rod-rod interactions,
with the equation of state for a  hard disk system to constrain the areal rod density.
The predicted nematic-membrane phase boundary shows reasonable agreement with simulation results,
and the predicted isolated-smectic phase boundary qualitatively agrees with simulation results for rod length $L<100\sigma$.
The predicted phase boundaries establish that there is a critical rod length $L\sim 30\sigma$,
below which isolated colloidal membrane can not be formed, for the given depletant size $\delta=1.5\sigma$.

The theoretical calculations enable systematic identification of the factors that control system phase behavior. In particular, the theoretical results demonstrate that correlations between protruding rods significantly enhance protrusion fluctuations and thereby profoundly affect membrane morphologies and the range of interactions between membranes. This effect is emphasized by comparison of the theoretically predicted protrusion distributions with those of a simpler theory that neglects correlations between neighboring rods. Evaluation of the theoretical predictions at different orders of the virial expansion further demonstrate that the terms up to third order that we have considered are essential for physical predictions. We speculate that including fourth order terms would lead to more quantitative agreement with the simulation results below the membrane crystallization point. At this level of sophistication, it will also be desirable to allow for rod orientational fluctuations, to determine their effect on the
locations of phase boundaries and to determine the effects of membrane bending modes on membrane-membrane interactions. As discussed earlier, these modes are expected to be of limited importance under the experimental conditions due to the high membrane bending modulus, but will become more important as the rod length is decreased. Finally, it would be useful to include effects of semi-flexibility. It was shown through scaling arguments in Ref. \cite{Yang2011} that semi-flexibility renormalizes the interactions between rods in the membrane leading to a smaller  equilibrium areal density; in addition, semiflexible rods will behave as if polydisperse in length.

{\bf Acknowledgement.} This work was supported by NSF-MRSEC-0820492 and NIH-R01AI080791. We thank Ed Barry and Zvonimir Dogic for many enlightening discussions. MFH gratefully acknowledges the support of KITP (where some of this manuscript was written) which is supported in part by the National Science Foundation under Grant No. NSF PHY05-51164.



\appendix
\section{Simulations of colloidal membranes}
\label{sec:simulations}
  In this section we describe the computer simulations whose results we have compared with the theory predictions. The simulations describe the equilibrium phase behavior for a model of hard rods and depletant molecules in the absence of any attractive interactions between rod ends and depletant \cite{Yang2011}.  The rods are represented as hard spherocylinders with diameter $\sigma$ and length $L$. The non-adsorbing polymer (depletant) is modeled with ghost spheres~\cite{Asakura1958} of diameter $\delta$, which act as hard spheres when interacting with rods but can freely interpenetrate one another. Compared with an effective pair potential approach \cite{Savenko2006,Patti2009,Cuetos2010}, this
model accounts for multi-rod interactions induced by polymers. Simulation results are reported with the following units: $\sigma$ is the unit of length, $k_BT$ is the unit of energy, and $k_BT\sigma^{-n}$ is the unit of $n$-dimensional pressure ($n = 2$ or $3$). As noted above, tilting of rods does not qualitatively affect membrane-membrane interactions and thus most free energy simulations were performed with rod orientations constrained parallel to the $z$ axis of the simulation box.

Examples of membrane-membrane interaction free energies calculated by umbrella sampling \cite{Frenkel2002} are shown in Fig. \ref{fig:freeEnergy} (bottom). For $\ps=0.06$ the free energy calculated under the orientational constraint is compared to the free energy calculated with the constraint relaxed. The phase behavior of the computational model system was predicted as a function of the depletant osmotic pressure $\ps$, the rod aspect ratio $L$, and the size of the depletant ghost spheres $\delta$, which corresponds to the polymer radius of gyration. Fig. \ref{fig:phasesTheory} (bottom) shows a cross-section of the phase diagram in terms of $\ps$ and $L$.

\section{Areal rod densities}
\label{sec:rho2dTheory}
As noted in the main text, the virial expansion Eq.~\ref{eq:freeEnergyOriginal} cannot accurately predict the areal densities of rods $\rhotwod$ within a membrane because areal densities are high under conditions for which membranes are stable, and membranes crystallize under high osmotic pressures.

We therefore independently obtain $\rhotwod$ by noting that, under the assumption that  rods are parallel to $z$ direction, the cross section of a colloidal membrane can be considered as a 2D system of hard disks. We ignore the small density of polymer that is actually inside the membrane, and the osmotic pressure $\ps$ determines the two-dimensional pressure felt by the hard disks (rods). The equation of state for hard disk systems has been studied extensively\cite{Luding2001,Brunner2003,mak2006,Mulero2008}.
Here we use the global equation of state given by Luding \cite{Luding2001} which characterizes both the liquid and crystalline  phases.
The equation of state is given by
\begin{align}
\beta p_\text{2d}/\rhotwod -1 = P_4 + m(\nu)(P_\text{dense}-P_4)
\label{eq:state2d}
\end{align}
where $\nu=\rhotwod\pi\sigma^2/4$ is the areal fraction. $P_4$ is the low density result
\begin{align}
P_4=2\nu g_4(\nu)
\end{align}
with
\begin{align}
g_4(\nu)=\frac{1-7\nu/16}{(1-\nu)^2}-\frac{\nu^3/16}{8(1-\nu)^4}
\end{align}
and $P_\text{dense}$ is the high density result
\begin{align}
P_\text{dense}=\frac{c_0}{\nu_\text{max}-\nu}h_3(\nu_\text{max}-\nu)-1
\label{eq:Cfour}
\end{align}
with $\nu_\text{max}=\pi/(2\sqrt{3})$ the maximum areal fraction, $h_3(x)=1+c_1x+c_3x^3$ a fit polynomial, and constants $c_0=1.8137$, $c_1=-0.04$ and $c_3=3.25$.
 Full details are in \cite{Luding2001}.
Eq.~\ref{eq:state2d} gives the pressure $p_\text{2d}$ as a function of the density,
but the density can be numerically inverted for a given pressure to give:
\begin{align}
\rhotwod&=\rhotwod(p_\text{2d})
\label{eq:rhotwod}
\end{align}

The 2D pressure $p_\text{2d}$ is the result of polymer osmotic pressure acting laterally to the membrane.
The excluded volume per rod in the membrane is approximately $v_\text{ex}\approx (L+h)/\rhotwod$, where we neglect rod protrusions.
Mechanical equilibrium then requires the 2D pressure to be
\begin{align}
p_\text{2d}=-\rho_\text{2d}^2\frac{\partial v_\text{ex}}{\partial \rhotwod}\ps\approx (L+h)\ps
\label{eq:ptwod}
\end{align}
The areal density of rods $\rhotwod(L,h,\ps)$ is then acquired from Eq.~\ref{eq:rhotwod} and Eq.~\ref{eq:ptwod} for a given $L$, $h$, and $\ps$.

Predicted values of $\rhotwod$ are compared to simulation results in Fig.~\ref{fig:rho2d}.
The simulation results of $\rhotwod$ are obtained from stable membranes of 256 rods; areal densities are not sensitive to system size.
 We find that the difference between simulated and predicted areal densities is within $5\%$, and estimated values are always lower than the simulation results.
The difference can be attributed to the rough estimation of the excluded volume per rod and the neglect of rod protrusions, as well as the accuracy of Eq.~\ref{eq:state2d}\cite{Luding2001}.
The sharp increase of $\rhotwod$, starting near $\rhotwod\approx 0.88$,
 identifies the transition from liquid phase to crystal phase (identified from the two-dimensional radial distribution function of rods $g(r)$ \cite{Yang2011}), with $\rhotwod\approx 0.88$ the hard disk freezing point \cite{Lowen1993,Truskett1998}.

\section{Integration of free energy}
\label{sec:integration}
Because rods are required to be parallel to $z$ direction, the free energy expression Eq.~\ref{eq:freeEnergyOriginal} can be greatly simplified.
The rod-rod Mayer function $f(\mathbf 1, \mathbf 2)\equiv \exp(-\beta U(\mathbf 1, \mathbf 2)) - 1$ can be separated into to the Mayer function
in $z$ direction and the Mayer function in $x-y$ plane\cite{Mulder1987},
\begin{align}
f(\mathbf 1, \mathbf 2)=-f_z(\mathbf 1, \mathbf 2)f_{xy}(\mathbf 1, \mathbf 2)
\end{align}
with
\begin{align}
f_z(\mathbf 1, \mathbf 2)=&\begin{cases}-1, &|z_1-z_2|<L\\0,&\text{otherwise}\end{cases}\nonumber\\
f_{xy}(\mathbf 1, \mathbf 2)=&\begin{cases}-1, &(x_1-x_2)^2+(y_1-y_2)^2<\sigma^2\\0,&\text{otherwise}\end{cases}
\end{align}
Since the rod density depends only on $z$, the integrations in Eq.~\ref{eq:freeEnergyOriginal} can be separated as well. We have,
\begin{align}
\beta f=&\int dz\rho(z)(\ln\rho(z) -1)\nonumber\\
&-\frac{1}{2}A\int\left(\prod_{i=1}^2 dz_i\rho(z_i)\right)f_z(\mathbf 1,\mathbf 2)\nonumber\\
&-\frac{1}{6}B\int \left(\prod_{i=1}^3 dz_i\rho(z_i)\right)f_z(\mathbf 1,\mathbf 2)f_z(\mathbf 1,\mathbf 3)f_z(\mathbf 2,\mathbf 3)\nonumber\\
&+\beta \ps\int dz(1-\alpha(z))
\label{eq:freeEnergyIntermediate}
\end{align}
with
\begin{align}
A\equiv&-\frac{1}{S_{xy}}\int\left(\prod_{i=1}^2dx_idy_i\right)f_{xy}(\mathbf 1, \mathbf 2)\nonumber\\
 =&\pi\sigma^2
\label{eq:A}
\end{align}
and
\begin{align}
B\equiv&-\frac{1}{S_{xy}}\int \left(\prod_{i=1}^3dx_idy_i\right)f_{xy}(\mathbf 1, \mathbf 2)f_{xy}(\mathbf 1, \mathbf 3)f_{xy}(\mathbf 2, \mathbf 3)\nonumber\\
=&\sigma^4\left(\pi^2-\frac{3\sqrt{3}}{4}\pi\right)
\end{align}

The integrations over $z$ can be further simplified,
\begin{align}
\int&\left(\prod_{i=1}^2 dz_i\rho(z_i)\right)f_z(\mathbf 1,\mathbf 2)\nonumber\\
&=-\int dz_1\rho(z_1)\int_{z_1-L}^{z_1+L} dz_2\rho(z_2)\nonumber\\
&=-\int dz_1\rho(z_1)(\rho^\dagger(z_1+L/2)+\rho^\dagger(z_1-L/2))
\end{align}
and
\begin{align}
\int&\left(\prod_{i=1}^3 dz_i\rho(z_i)\right)f_z(\mathbf 1,\mathbf 2)f_z(\mathbf 1,\mathbf 3)f_z(\mathbf 2,\mathbf 3)\nonumber\\
&=-\int dz_1\rho(z_1)\int_{z_1-L}^{z_1}dz_2 \rho(z_2) \int_{z_2}^{z_2+L}dz_3 \rho(z_3)\nonumber\\
&\quad - (z_2\Leftrightarrow z_3) \nonumber\\
&\quad-\int dz_1\rho(z_1)\int_{z_1}^{z_1+L}dz_2\rho(z_2) \int_{z_1}^{z_1+L}dz_3\rho(z_3)\nonumber\\
&=-2\int dz_1 \rho(z_1)\int_{z_1-L}^{z_1}dz_2\rho(z_2)\rho^\dagger(z_2+L/2)\nonumber\\
&\quad -\int dz_1\rho(z_1)(\rho^\dagger(z_1+L/2))^2
\label{eq:virialIntegration}
\end{align}
with $\rho^\dagger(z)$ defined in Eq.~\ref{eq:rhoDagger}.
The final expression for the free energy, Eq.~\ref{eq:freeEnergy}, is then acquired by substituting equations from Eq.~\ref{eq:A} to Eq.~\ref{eq:virialIntegration} into Eq.~\ref{eq:freeEnergyIntermediate}.

\section{Free energy minimization}
\label{sec:differentiation}
The two terms in Eq.~\ref{eq:fmin} are calculated as,
\begin{align}
\frac{\delta \beta \fff}{\delta \rho(z)} =& \ln\rho(z) + A(\rho^\dagger(z+L/2)+\rho^\dagger(z-L/2))\nonumber\\
&+B\int_{z-L}^z dz_2\rho(z_2)\rho^\dagger(z_2+L/2) \nonumber\\
&+\frac{1}{2}B(\rho^\dagger(z_1+L/2))^2\nonumber\\
&-\beta\ps\int dz_1\frac{\partial \alpha(z_1)}{\partial \rho(z)}
\end{align}
and
\begin{align}
\frac{\delta C}{\delta \rho(z)} = 1
\end{align}
Note that the factor $1/2$ in front of $A$ and the factor $1/3$ in front of $B$ are canceled because $\rho$ appears multiple times in the corresponding terms (see Eq.~\ref{eq:freeEnergyOriginal}). Substituting these results into Eq.~\ref{eq:fmin}, the resulting integral equation of $\rho$ is acquired as Eq.~\ref{eq:multiLayerRho}. When $\rho$ is numerically solved, the constraint Eq.~\ref{eq:constraint} is applied through $\zeta$ in each iteration step.




\bibliographystyle{apsrev4-1}
\bibliography{ref.clean}

\end{document}